\documentclass[prl,twocolumn,superscriptaddress]{revtex4-2}
\usepackage[dvipdfmx]{graphicx}
\usepackage{type1cm}
\usepackage[usenames,svgnames]{xcolor}
\usepackage{color}
\usepackage{url}
\usepackage{natbib}
\usepackage[colorlinks=true,linkcolor=blue,citecolor=blue,breaklinks=true]{hyperref}

\usepackage{amsmath,amssymb,bm,amsthm}
\usepackage{libertine}
\usepackage[libertine]{newtxmath}

\DeclareMathOperator{\Tr}{\mathrm{Tr}}
\DeclareMathOperator{\re}{\mathrm{Re}}
\DeclareMathOperator{\im}{\mathrm{Im}}
\newcommand{\dd}{d}

\newcommand{\dv}[1]{\frac{d}{d#1}}

\usepackage{braket}

\newcommand{\calL}{\mathcal{L}}

\usepackage[capitalize]{cleveref}

\newcommand{\sectionprl}[1]{{\par\it #1.---}}

\begin{document}

\title{Accelerated Decay due to Operator Spreading in Bulk-Dissipated Quantum Systems}

\author{Tatsuhiko Shirai}
\affiliation{Waseda Institute for Advanced Study, Waseda University, Nishi Waseda, Shinjuku-ku, Tokyo 169-0051, Japan}

\author{Takashi Mori}
\affiliation{Department of Physics, Keio University, Kohoku-ku, Yokohama, Kanagawa 223-8522, Japan}

\begin{abstract}
Markovian open many-body quantum systems display complicated relaxation dynamics.
The spectral gap of the Liouvillian characterizes the asymptotic decay rate towards the stationary state, but it has recently been pointed out that the spectral gap does not necessarily determine the overall relaxation time.
Our understanding on the relaxation process before the asymptotically long-time regime is still limited.
We here present a collective relaxation dynamics of autocorrelation functions in the stationary state.
As a key quantity in the analysis, we introduce the instantaneous decay rate, which characterizes the transient relaxation and converges to the conventional asymptotic decay rate in the long-time limit.
Our theory predicts that a bulk-dissipated system generically shows an accelerated decay before the asymptotic regime due to the scrambling of quantum information associated with the operator spreading.
\end{abstract}
\maketitle

\sectionprl{Introduction}
It is a fundamental problem in nonequilibrium statistical physics to elucidate how a quantum system approaches to stationality under dissipative couplings to large environments~\citep{breuer2002theory, spohn1980kinetic, znidaric2015relaxation}.
This problem has been investigated mainly for small quantum systems, whereas our understanding on the nonequilibrium dynamics of open \emph{many-body} quantum systems has still been limited.
The recent experimental progress using ultracold atoms and trapped ions has made us to introduce controlled dissipation to realize many-body quantum systems with desired properties~\citep{diehl2008quantum, verstraete2009quantum, barreiro2011open, barontini2013controlling, tomita2017observation}.
The experimental background also motivates us to study the generic dynamical properties of open many-body quantum systems.


Recently, it was pointed out that open many-body quantum systems exhibit counterintuitive dynamical features.
The dynamics of Markovian open quantum systems are generated by the Liouvillian superoperator of the celebrated Lindblad form~\citep{lindblad1976generators, gorini1976completely}.
The spectral gap of the Liouvillian, which is simply called the Liouvillian gap, gives the asymptotic decay rate~\citep{znidaric2015relaxation}, and thus it is expected that the relaxation time is given by the inverse of the Liouvillian gap.
However, it was found that the relaxation time can be much longer than the inverse of the Liouvillian gap when there is a conserved current in the bulk (it is always the case when the dissipation acts only at the boundaries of the system)~\citep{mori2020resolving, haga2021liouvillian, lee2023anomalously, sawada2024role}.
The point is that the crossover time into the asymptotic regime is very long and even diverging in the thermodynamic limit.
If the transient regime is dominant for the overall relaxation process, the decay rate in the transient regime determines the relaxation time.
Hence, we should investigate the relaxation dynamics in the transient regime, which is the main focus of our work.



\begin{figure}
    \centering
    \includegraphics[width=1\linewidth]{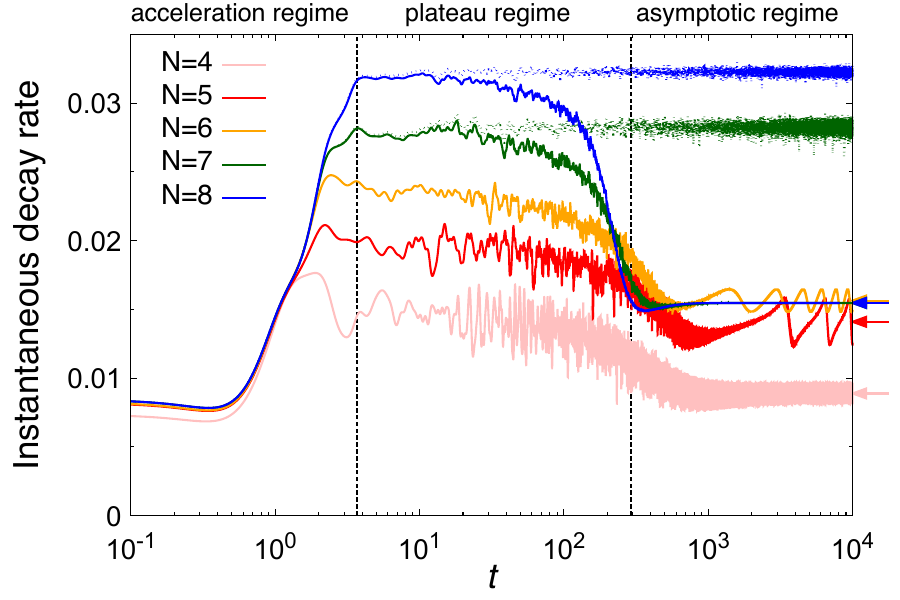}
    \caption{Dynamics of the instantaneous decay rate $\kappa_A(t)$ with $\hat{A}=\sigma_1^z-\braket{\sigma_1^z}_{\rm ss}$ for various system sizes in the bulk-dissipated Ising spin chain (see the model in \cref{Eq:system_Hamiltonian} and \cref{Eq:jump_operator}).
    Solid lines show $\kappa_A(t)$ for $N=4,5,6,7,8$ from bottom to top, while dotted lines show $\kappa_A^{(0)}(t)$ for $N=7,8$.
    The arrows indicate the values of the Liouvillian gap.
    }
    \label{fig:instance_gap}
\end{figure}

In this Letter, we discuss the transient dynamics of quantum many-body systems under \emph{bulk dissipation}.
In contrast to boundary dissipation, we find that the relaxation time is generically much \emph{shorter} than the inverse of the Liouvillian gap in bulk-dissipated systems.
Based on a rigorous inequality on the autocorrelation function, we introduce a key quantity called the instantaneous decay rate.
We argue that the instantaneous decay rate exhibits three distinct dynamical regimes in the weak dissipation regime: (1) the acceleration regime, (2) the plateau regime, and (3) the asymptotic regime.
\Cref{fig:instance_gap} illustrates dynamics of the instantaneous decay rate for various system sizes in a bulk-dissipated system.
In the acceleration regime, the decay rate increases with time.
This is a universal phenomenon under bulk dissipation, and this accelerated decay is explained by the operator-size growth (or the operator spreading) under unitary time evolution without dissipation.
In the plateau regime, the decay rate is saturated at a constant value proportional to $\gamma N$, where $\gamma$ and $N$ are the strength of dissipation and the system size, respectively.
In the asymptotic regime, the decay rate converges to the asymptotic value, which is identical to the Liouvillian gap.
In the following, we first provide the general set up, and then present the main result.
We demonstrate the relevance of the theoretical results in a bulk-dissipated spin chain.

\sectionprl{Set up}
This work adopts the same set up as \citep{mori2023symmetrized}.
We consider a finite quantum system, whose density matrix $\rho(t)$ obeys the Lindblad equation~\citep{breuer2002theory}:
\begin{align}
    &\frac{\dd\rho(t)}{\dd t}=\calL\rho(t),\nonumber\\
    &\calL\rho=-i[\hat{H},\rho]+\gamma \sum_k  \left(\hat{L}_k\rho\hat{L}_k^\dagger-\frac{1}{2}\left\{\hat{L}_k^\dagger\hat{L}_k,\rho\right\}\right),
    \label{eq:Liouvillian}
\end{align}
where $\calL$ is the Liouvillian superoperator, $\hat{H}$ is the bulk Hamiltonian, and $\{\hat{L}_k\}$ is a set of jump operators.
The Planck constant $\hbar$ is set to unity.
We assume that the stationary state is unique.
The density matrix of the stationary state $\rho_\mathrm{ss}$ is described by $\calL \rho_\mathrm{ss} = 0$.
The Lindblad equation ensures the complete positivity~\citep{lindblad1976generators, gorini1976completely}.

Let $\hat{A}$ be an Hermitian operator.
The autocorrelation function for $\hat{A}$ in the stationary state is expressed as~\citep{carmichael1998statistical}
\begin{align}
    C_{AA}(t) \coloneqq \braket{\hat{A}(t), \hat{A}}_\mathrm{ss},
\end{align}
where $\braket{\hat{A}, \hat{B}}_\mathrm{ss} = \Tr(\hat{A}^\dagger\hat{B} \rho_\mathrm{ss})$.
We can assume $\braket{\hat{A}}_\mathrm{ss} \coloneqq \Tr (\hat{A} \rho_\mathrm{ss})=0$ without loss of generality.
Here, $\hat{A}(t)$ represents the time-evolved operator in the Heisenberg picture: $\hat{A}(t)=e^{\tilde{\calL} t}\hat{A}$, where the superoperator $\tilde{\calL}$ is defined as
\begin{align}
    \tilde{\calL} \hat{A} = i[\hat{H},\hat{A}] +\gamma \sum_k\left(\hat{L}_k^\dagger\hat{A}\hat{L}_k-\frac{1}{2} \left\{\hat{L}_k^\dagger\hat{L}_k,\hat{A}\right\}\right).
\end{align}
The eigenvalue spectrum of $\tilde{\calL}$ is the same as that of $\calL$.
Since any eigenvalue has a non-positive real part, we sort the eigenvalues denoted by $\{\lambda_\alpha\}$ in the descending order: $0=\lambda_0 > \re\lambda_1 \geq \re\lambda_2 \geq \dots$.
The Liouvillian gap $g$ is defined as
\begin{align}
    g=-\re\lambda_1.
\end{align}
The Liouvillian gap determines the asymptotic decay rate of $C_{AA}(t)$ as
\begin{equation}
    |C_{AA}(t)| \sim e^{-g t} \quad (t \to \infty).
\end{equation}

In this work, we discuss the transient dynamics before reaching the asymptotic regime.
Recent work introduced the symmetrized Liouvillian to study the relaxation dynamics in a transient regime~\citep{mori2023symmetrized}:
\begin{align}
    \tilde{\mathcal{L}}_s=\frac{\tilde{\mathcal{L}}+\tilde{\mathcal{L}}^*}{2},
\end{align}
where the superoperator $\tilde{\mathcal{L}}^*$ is the adjoint superoperator of $\tilde{\mathcal{L}}$ associated with the inner product $\braket{\hat{A}, \hat{B}}_\mathrm{ss}$:
\begin{align}
\braket{\hat{A},\tilde{\calL}\hat{B}}_\mathrm{ss}=\braket{\tilde{\calL}^*\hat{A},\hat{B}}_\mathrm{ss},
\end{align}
which is expressed as $\tilde{\calL}^* \hat{A}=\calL(\hat{A}\rho_\mathrm{ss})\rho_\mathrm{ss}^{-1}$~\citep{alicki1976detailed}.
We assume that $\rho_\mathrm{ss}$ is invertible for simplicity.
The symmetrized Liouvillian is a non-positive and Hermitian superoperator (i.e., $\tilde{\mathcal{L}}_s \leq 0$ and $\tilde{\mathcal{L}}_s^* = \tilde{\mathcal{L}}_s$).
The non-positivity is shown by using the complete positivity of $\mathcal{L}$.
We sort the eigenvalues of $\tilde{\mathcal{L}}_s$ denoted by $\{ s_{\alpha} \}$ in the descending order: $0=s_0 \geq s_1 \geq s_2\geq\dots$.
$\tilde{\mathcal{L}}_s$ has a zero eigenvalue: $\tilde{\mathcal{L}}_s \hat{I} = 0$, where $\hat{I}$ is the identity operator.
The spectral gap $g_s$ of $\tilde{\mathcal{L}}_s$ is defined by
\begin{align}
g_s = -s_1.
\end{align}
The spectral gap gives a simple bound on the autocorrelation function in the stationary state~\citep{mori2023symmetrized}:
\begin{align}
    |C_{AA}(t)|\leq e^{-g_s t}C_{AA}(0).
    \label{eq:bound_gap}
\end{align}
However, as we see in the following, we need a more sophisticated analysis beyond the spectral gap bound like \cref{eq:bound_gap} to describe generic relaxation dynamics of bulk-dissipated quantum many-body systems.

\sectionprl{Instantaneous decay rate}
\begin{figure}
    \centering
    \includegraphics[width=1\linewidth]{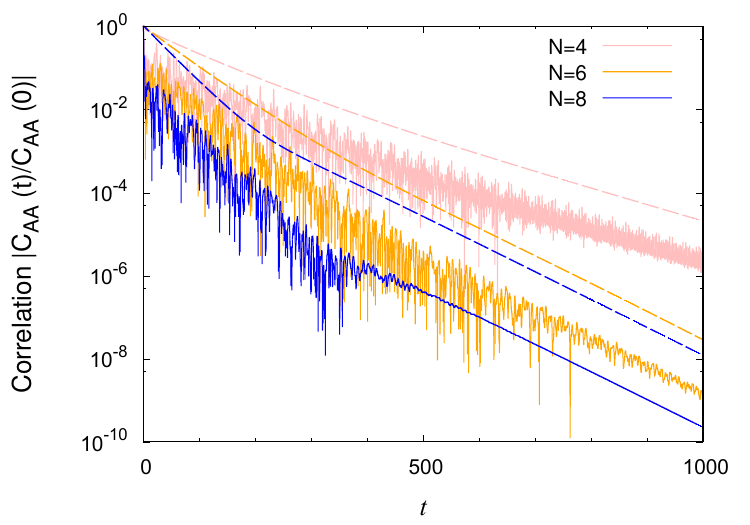}
    \caption{Dynamics of the autocorrelation function $C_{AA}(t)$ with $\hat{A}=\hat{\sigma}_1^z - \braket{\hat{\sigma}_1^z}_{\rm ss}$ for various system sizes in the bulk-dissipated Ising spin chain.
    Solid lines show $|C_{AA}(t)/C_{AA}(0)|$ for $N=4, 6, 8$ from top to bottom.
    Dashed lines show the upper bound of the correlation functions in \cref{eq:C_inst}.
    Although the upper bounds are not tight, they correctly capture large decay rates appearing in the transient regime.}
    \label{fig:correlation_decay}
\end{figure}
Let us investigate how the autocorrelation function in the stationary state decays.
Although the discussion below is general, for clarity we consider a bulk-dissipated Ising spin chain under the periodic boundary condition \citep{supp}.
The bulk Hamiltonian is given by
\begin{align}
    \hat{H} = \sum_{i=1}^N \left( h^z \hat{\sigma}_i^z + h^x \hat{\sigma}_i^x + J \hat{\sigma}_i^z \hat{\sigma}_{i+1}^z \right),
    \label{Eq:system_Hamiltonian}
\end{align}
where $\hat{\sigma}_i^{x,y,z}$ denotes the Pauli matrices at site $i$.
We fix the parameters as $h^z=0.9045$, $h^x=0.809$, and $J=1$, for which eigenstate thermalization hypothesis has been numerically shown to hold \citep{kim2014testing}.
Each site $i$ is associated with the jump operator $\hat{L}_i$ given by
\begin{align}
    \hat{L}_i = \hat{\sigma}_i^- = \frac{1}{2} (\hat{\sigma}_i^x - i \hat{\sigma}_i^y).
    \label{Eq:jump_operator}
\end{align}
We assume weak dissipation and set $\gamma = 0.01$.
This open quantum system has been implemented on Rydberg atoms under laser driving and dissipation~\citep{carr2013nonequilibrium, letscher2017bistability}.
Each Rydberg atom is regarded as a two-level system, where the spin up (down) corresponds to the Rydberg (ground) state.
In this model, the Liouvillian gap is independent of the system size (i.e., $g \sim O(N^0)$. See \cref{fig:instance_gap}.).

Figure~\ref{fig:correlation_decay} shows the autocorrelation functions of a local operator $\hat{A}=\hat{\sigma}_1^z - \braket{\hat{\sigma}_1^z}_{\rm ss}$ for various system sizes by solid lines.
In the long-time limit, the autocorrelation function decays at the rate of the Liouvillian gap.
There is a different decay rate appearing in the transient regime.
The decay rate is much larger than the Liouvillian gap.

To understand how the transient dynamics emerges, we give an upper bound of the autocorrelation function:
\begin{align}
    |C_{AA}(t)|\leq e^{-\int_0^t\dd s\, \kappa_A(s)} C_{AA}(0).
    \label{eq:C_inst}
\end{align}
Here, we introduce the \emph{instantaneous decay rate} $\kappa_A(t)$.
The instantaneous decay rate is defined by using the symmetrized Liouvillian $\tilde{\mathcal{L}}_s$ as
\begin{align}
    \kappa_A(t)=-\frac{\braket{\hat{A}(t),\tilde{\mathcal{L}}_s\hat{A}(t)}_\mathrm{ss}}{\braket{\hat{A}(t),\hat{A}(t)}_\mathrm{ss}}.
    \label{eq:inst_decay}
\end{align}

We prove the inequality.
The autocorrelation function is generally bounded as follows:
\begin{align}
    |C_{AA}(t)|\leq\sqrt{\braket{\hat{A}(t),\hat{A}(t)}_\mathrm{ss}}\sqrt{\braket{\hat{A},\hat{A}}_\mathrm{ss}},
    \label{eq:bound_C}
\end{align}
where we have used the Cauchy-Schwarz inequality $|\braket{\hat{A},\hat{B}}_\mathrm{ss}| \leq \sqrt{\braket{\hat{A},\hat{A}}_\mathrm{ss}\braket{\hat{B},\hat{B}}_\mathrm{ss}}$.
The time evolution of the quantity $\braket{\hat{A}(t),\hat{A}(t)}_\mathrm{ss}$ is given by
\begin{align}
    \dv{t}\braket{\hat{A}(t),\hat{A}(t)}_\mathrm{ss}=2\braket{\hat{A}(t),\tilde{\mathcal{L}}_s\hat{A}(t)}_\mathrm{ss}.
    \label{eq:EOM_AA}
\end{align}
\Cref{eq:EOM_AA} is then formally solved as
\begin{align}
    \braket{\hat{A}(t),\hat{A}(t)}_\mathrm{ss}=e^{-2\int_0^t\dd s\,\kappa_A(s)}\braket{\hat{A},\hat{A}}_\mathrm{ss}.
\end{align}
By substituting it into \cref{eq:bound_C}, we obtain \cref{eq:C_inst}.
The instantaneous decay rate is actually the decay rate of the quantity $\sqrt{\braket{\hat{A}(t),\hat{A}(t)}_\mathrm{ss}}$, and it gives an upper bound on the autocorrelation function as in \cref{eq:C_inst}.

Now we list important properties of the instantaneous decay rate.
Firstly, it is non-negative $\kappa_A(t)\geq 0$ since $\tilde{\mathcal{L}}_s$ is a non-positive superoperator.
Secondly, $\kappa_A(t)$ is not smaller than the spectral gap $g_s$ of $\tilde{\mathcal{L}}_s$:
\begin{align}
    \kappa_A(t)\geq g_s.
    \label{eq:property_gap_K}
\end{align}
This is proved by using the following property:
\begin{align}
    g_s = \inf_{\hat{X}\neq 0, \braket{\hat{X}}_\mathrm{ss}=0}\frac{\braket{\hat{X},\tilde{\mathcal{L}}_s\hat{X}}_\mathrm{ss}}{\braket{\hat{X},\hat{X}}_\mathrm{ss}}.
    \label{eq:gap_K}
\end{align}
The condition $\braket{\hat{X}}_\mathrm{ss}=\braket{\hat{I},\hat{X}}_\mathrm{ss}=0$ ensures that $\hat{X}$ is orthogonal to $\hat{I}$, which is the eigenvector of $\tilde{\mathcal{L}}_s$ corresponding to the zero eigenvalue, and thus $g_s$ is expressed in the variational form as in \cref{eq:gap_K}.
\Cref{eq:property_gap_K} indicates that the instantaneous decay rate gives a simple bound on the autocorrelation function, i.e. Eq.~(\ref{eq:bound_gap}), but it cannot capture the three dynamic regimes manifest in Fig.~\ref{fig:instance_gap}.


Thirdly, the instantaneous decay rate converges to the asymptotic decay rate (i.e. the Liouvillian gap) in the long-time limit:
\begin{align}
    \lim_{t\to\infty}\kappa_A(t)=g,
\end{align}
when $\Tr (\hat{A} \rho_1) \neq 0$ and $\im \lambda_n = 0$ for every $n$ with $\re \lambda_n = -g$.

\sectionprl{Accelerated decays due to operator spreading}
To gain insights into the relaxation dynamics in \cref{fig:correlation_decay}, we investigate how $\kappa_A(t)$ behaves at transient times for sufficiently small $\gamma$ with a fixed system size $N$.

\Cref{fig:instance_gap} illustrates the dynamics of $\kappa_A(t)$ by solid lines.
We find three distinct dynamic regimes: acceleration regime, plateau regime, and asymptotic regime.
$\kappa_A(t)$ initially increases in the acceleration regime, has an almost constant value in the plateau regime, and decreases and converges to the Liouvillian gap in the asymptotic regime.
The growth of $\kappa_A(t)$ implies the acceleration of the decay.

For weak dissipation, it is expected that the time evolution of $\hat{A}(t)$ appearing in the instantaneous decay rate, i.e. \cref{eq:inst_decay}, would be well approximated by the unitary dynamics without dissipation.
The instantaneous decay rate is thus approximated as
\begin{align}
    \kappa_A^{(0)}(t) = -\frac{\braket{\hat{A}_0(t),\tilde{\mathcal{L}}_s\hat{A}_0(t)}_\mathrm{ss}}{\braket{\hat{A}_0(t),\hat{A}_0(t)}_\mathrm{ss}}, \quad \hat{A}_0(t) = e^{i\hat{H}t}\hat{A}e^{-i\hat{H}t}.
    \label{eq:approx}
\end{align}
Figure~\ref{fig:instance_gap} shows the dynamics of $\kappa_A^{(0)}(t)$ for $N=7$ and $8$.
The approximated one correctly reproduces the instantaneous decay rate in the acceleration regime and the plateau regime~\footnote{Note that there is a small deviation between $\kappa_A(t)$ and $\kappa_A^{(0)}(t)$ within the plateau regime, which linearly increases over time.}. 
It indicates that the growth of the instantaneous decay rate arises due to the unitary dynamics and dissipation is relevant only for the crossover to the asymptotic regime.

\begin{figure}
    \centering
    \includegraphics[width=1\linewidth]{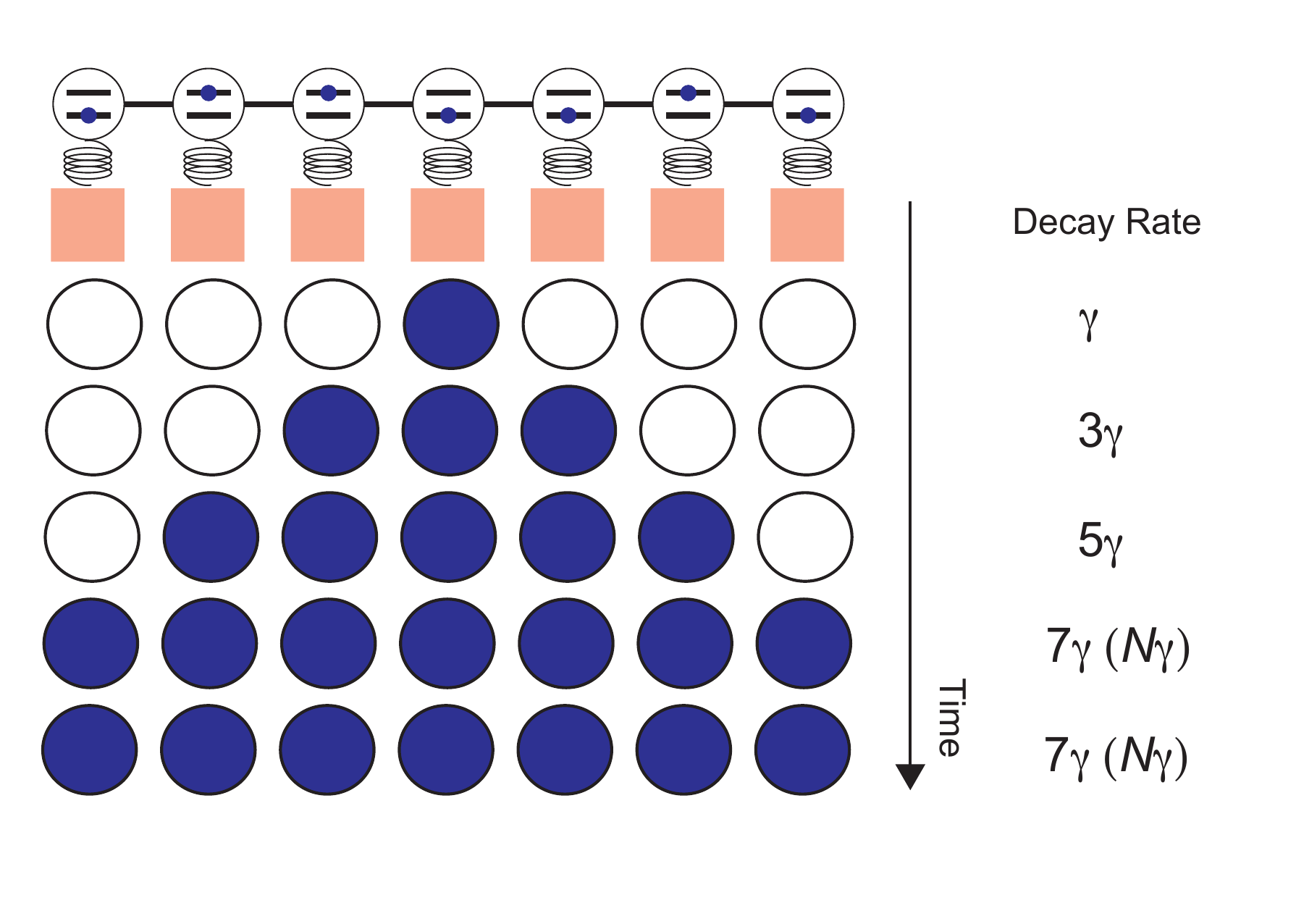}
    \caption{Mechanism of accelerated decay in a bulk-dissipated quantum chain with $N=7$ spins.
    The vertical axis denotes the time.
    Colored circles represent sites where an operator $\hat{A}^{(0)}(t)$ nontrivially acts.
    }
    \label{fig:schematic}
\end{figure}

We now explain the growth of the instantaneous decay rate from the viewpoint of operator spreading.
\Cref{fig:schematic} gives a schematic of operator spreading in a one-dimensional quantum spin system.
Let us consider a local operator $\hat{A}$ acting to a single site at initial time.
In the figure, colored circles denote the sites where the operator $\hat{A}_0(t)$ nontrivially acts.
Under unitary dynamics, the operator spreads over the system and the number of colored sites linearly increases with time \citep{keyserlingk2018operator, nahum2018operator}.
Here, the number of colored sites represents the operator size.
Then, we consider the effect of bulk dissipation.
The instantaneous decay rate is roughly estimated as the product of $\gamma$ and the operator size.
Suppose an operator $\hat{O}_i$ acting on a single site $i$.
The operator decays with a decay rate of $\gamma$ since the dissipation independently acts on every site with strength $\gamma$.
Then, the operator acting to three sites such as $\hat{O}_{i-1}\hat{O}_{i}\hat{O}_{i+1}$ decays with a decay rate of $3\gamma$.
In this way, the instantaneous decay rate accelerates due to the operator spreading in the acceleration regime.

In the plateau regime, the instantaneous decay rate is saturated to a value proportional to $N$.
The plateau regime is also explained from operator spreading.
In this regime, the operator $\hat{A}^{(0)}(t)$ spreads across the entire system, and hence the operator size is $N$.
Using the relation between the operator size and the instantaneous decay rate, $\kappa_A(t) \sim \gamma N$.
The dissipation exhibits a collective decay in this regime.

In the asymptotic regime, the instantaneous decay rate converges to the Liouvillian gap $g$.
The oscillation around the Liouvillian gap in \cref{fig:instance_gap} implies that $\im \lambda_1 \neq 0$.
We find in supplemental material that the crossover time into the asymptotic regime is inversely proportional to $\gamma$ for weak dissipation \citep{supp}.
Thus, the plateau regime is longer for weaker dissipation.

We develop a more quantitative theory explaining those three dynamical regimes.
In the following argument, we assume $\gamma \ll N^{-1}$, which is relevant for this study.
Later, we will briefly discuss what happens when $N\to \infty$ is taken at a fixed $\gamma$.
The instantaneous decay rate $\kappa_A(t)$ obeys the following equation \citep{supp}:
\begin{equation}
    \frac{\dd \kappa_A(t)}{\dd t}= v_A -2 \delta \kappa_A^2,
    \label{eq:time-derivative}
\end{equation}
where
\begin{align}
    v_A &= \frac{\braket{\hat{A}(t),[\tilde{\mathcal{L}}, \tilde{\mathcal{L}}_s] \hat{A}(t)}_\mathrm{ss}}{\braket{\hat{A}(t),\hat{A}(t)}_\mathrm{ss}}, \nonumber \\
    \delta \kappa_A^2 &= \frac{\braket{\hat{A}(t), (-\tilde{\mathcal{L}}_s - \kappa_A(t))^2 \hat{A}(t)}_\mathrm{ss}}{\braket{\hat{A}(t),\hat{A}(t)}_\mathrm{ss}}.
\end{align}
This formula generalizes the result in~\citep{schuster2023operator} to any open quantum systems obeying Lindblad equations.
$v_A / \gamma$ represents how fast the operator size of $A(t)$ spreads with time and $(\delta \kappa_A /\gamma)^2$ roughly gives the variance of the operator size.
Note that the time evolution of $A(t)$ in $v_A$ is well approximated by the unitary dynamics without dissipation~\citep{supp}.

The formula accurately explains the observed three dynamic regimes in \cref{fig:instance_gap}.
In the acceleration regime, recent studies on random unitary circuits~\citep{keyserlingk2018operator,nahum2018operator} indicate $v_A \sim \gamma$ and $\delta \kappa_A^2 \sim \gamma^2 t$.
Since $v_A \gg \delta \kappa_A^2$ for small $\gamma$, 
\begin{equation}
    \kappa_A(t) \sim \gamma t \quad (t \lesssim N).
\end{equation}
The system enters the plateau regime when the operator spreads across the entire system at $t \sim N$.
Accordingly, $\kappa_A(t)$ is proportional to $\gamma N$.
In this regime, $v_A \approx 0$, whereas $\delta \kappa_A^2 \sim \gamma^2 N$~\citep{supp}.
Thus, during the duration time $\Delta t$ in the plateau regime, $\kappa_A(t)$ reduces by the amount proportional to $\gamma^2 N \Delta t$.
This decrease is negligible compared with the plateau value $\sim \gamma N$ up to $\Delta t \sim \gamma^{-1}$.
We therefore conclude that $\kappa_A(t)$ approximately remains constant up to $t \sim \gamma^{-1}$:
\begin{equation}
    \kappa_A(t) \sim \gamma N \quad (N \lesssim t \lesssim \gamma^{-1}).
\end{equation}
When $t \gtrsim \gamma^{-1}$, the variation of $\kappa_A(t)$ becomes the same order of magnitude as the plateau value.
Then the system shows a crossover to the asymptotic regime and $\kappa_A(t)$ converges to the Liouvillian gap.

Here, we briefly explain the dynamics of $\kappa_A(t)$ in the thermodynamic limit (i.e., $N \to \infty$ before taking $\gamma \ll 1$).
\Cref{eq:time-derivative} with $v_A \sim \gamma$ and $\delta \kappa_A^2 \sim \gamma^2 t$ implies that the operator-size growth stops at $t \sim \gamma^{-1}$ due to dissipation.
The peak of $\kappa_A(t)$ is independent of $\gamma$ and $N$.
Subsequently, $\kappa_A(t)$ gradually approaches the Liouvillian gap $g \sim \gamma$.
In this way, there is no plateau regime in the thermodynamic limit, but the amplification of dissipation via the operator spreading occurs, which yields a finite decay rate even for an infinitesimally small $\gamma$.

Finally we compare the exact autocorrelation function with its upper bound.
In \cref{fig:correlation_decay}, we plot the autocorrelation $|C_{AA}(t)/C_{AA}(0)|$ by solid lines, whereas the upper bound $e^{-\int_0^t ds \kappa_A(s)}$ by dashed lines.
The autocorrelation function shows a rapid decay due to the unitary dynamics at short times.
This effect is not taken into account in the upper bound.
Thus, we find that the inequality in \cref{eq:C_inst} is not tight.
However, the upper bound qualitatively reproduces the autocorrelation function at large times.
In particular, the relaxation dynamics in the transient regime is well described by the plateau value of the instantaneous decay rate.
The plateau value is proportional to the system size, which is consistent with the numerical observation that a larger system shows a faster decay.
Thus, the autocorrelation function for a local operator exhibits a collective decay due to the operator spreading.

\sectionprl{Summary and Discussion}
We studied the autocorrelation functions for a local operator in the stationary state in bulk-dissipated quantum many-body systems.
In the transient regime, the correlation exhibits a fast relaxation with a decay rate much larger than the asymptotic one.
We derived a rigorous upper bound on the autocorrelation function.
We demonstrated that the instantaneous decay rate shows a plateau over a long time interval (\cref{fig:instance_gap}), which explains the collective decay found in the autocorrelation function (\cref{fig:correlation_decay}).

We explained the mechanism of the accelerated decay from the operator spreading.
This mechanism is generic in quantum many-body systems under weak bulk dissipation.
In recent cold-atom experiments, bulk dissipation has been introduced in a controlled manner~\citep{tomita2017observation}.
Ultracold atoms thus provide an experimental platform to confirm the collective relaxations predicted in this work.
Also this set up is relevant to the field of quantum computation using noisy intermediate-scale quantum devices \citep{preskill2018quantum}.

As a comparison of decay rates indicated by $O(N)$,  we explain the difference with super-radiance.
Super-radiance is a phenomenon that occurs when $N$ atoms couple with a common dissipative environment~\citep{dicke1954coherence}.
Therefore, it is qualitatively different from the accelerated decay in this work, in which the dissipation independently acts on each site.

Although this work focuses on one-dimensional systems, the accelerated decay should also appear in high-dimensional systems.
Namely, the instantaneous decay rate increases as $\kappa_A(t)\sim \gamma t^d$ where $d$ is a spatial dimension in the acceleration regime, $\kappa_A(t) \sim \gamma N$ in the plateau regime, and $\kappa_A(t)=g$ in the asymptotic regime.
The crossover time between the plateau regime and the asymptotic regime should be longer for weaker dissipation.
It is a future problem to study the explicit $\gamma$ dependence of the crossover time in two or three dimensional systems.

\begin{acknowledgments}
This work was supported by JSPS KAKENHI (Grant Numbers JP21H05185 and 23K13034) and by JST, PRESTO Grant No. JPMJPR2259.
\end{acknowledgments}

\bibliographystyle{apsrev4-2}
\bibliography{apsrevcontrol,physics_v2}

\clearpage
\begin{widetext}

\begin{center}
\textbf{\Large Supplemental Material}

\bigskip
Tatsuhiko Shirai$^{1}$ and Takashi Mori$^{2}$\\
\textit{
${}^1$Waseda Institute for Advanced Study, Waseda University, Nishi Waseda, Shinjuku-ku, Tokyo 169-0051, Japan
}\\
\textit{
${}^2$Department of Physics, Keio University, Kohoku-ku, Yokohama, Kanagawa 223-8522, Japan
}

\end{center}

\setcounter{equation}{0}
\def\theequation{S\arabic{equation}}
\setcounter{figure}{0}
\def\thefigure{S\arabic{figure}}
\setcounter{secnumdepth}{2}
\renewcommand{\thesection}{\Alph{section}}

\makeatletter
\def\@hangfrom@section#1#2#3{\@hangfrom{#1#2}#3}
\def\@hangfroms@section#1#2{#1#2}
\makeatother

\section{Bose-Hubbard model}
To demonstrate the broad applicability of our results, we study a hard-core Bose-Hubbard model under bulk dissipation:
\begin{equation}
    \hat{H}=\sum_{i=1}^N \left[ -h \left( \hat{b}_i \hat{b}_{i+1}^\dagger +\hat{b}_i^\dagger \hat{b}_{i+1}\right) + V \hat{n}_i \hat{n}_{i+1} -h' \left( \hat{b}_i \hat{b}_{i+2}^\dagger +\hat{b}_i^\dagger \hat{b}_{i+2}\right) + V' \hat{n}_i \hat{n}_{i+2} \right] 
\end{equation}
and
\begin{equation}
    \hat{L}_i = \hat{n}_i  \text{ for } i=1,\ldots,N,
\end{equation}
where $\hat{b}_i$ and $\hat{b}_i^\dagger$ are annihilation and creation operators of hard-core bosons at site $i$, respectively. The eigenvalues of the operator $\hat{n}_i = \hat{b}_i^\dagger \hat{b}_i$ are $0$ and $1$.
The periodic boundary condition is assumed.
The parameters of the Hamiltonian are given by $(h, h', V, V')=(1,0.24,1,0.24)$, where the eigenstate thermalization hypothesis has been numerically shown \citep{rigol2009breakdown}.
The Lindblad operators describe a dephasing process and the strength is set to $\gamma=0.01$.
The half-filled sector (i.e., $\sum_{i=1}^N \hat{n}_i=N/2$) is considered, and the steady state has a uniform distribution in this sector.

\Cref{fig:BH} shows the dynamics of the instantaneous decay rate $\kappa_A (t)$ and the autocorrelation function $C_{AA}(t)$ with $\hat{A}=\hat{n}_1 - 1/2$.
As the spin-chain model in the main text, the instantaneous decay rate exhibits three dynamic regimes.
The autocorrelation function decays at a rate proportional to the system size, which corresponds to the plateau value of $\kappa_A(t)$ in \cref{fig:BH}(a).
It is noted that the crossover time between the plateau regime and the asymptotic regime is so large that we can not observe the asymptotic regime in \cref{fig:BH}(b).

\begin{figure*}[h]
    \centering
    \includegraphics[width=0.47\linewidth]{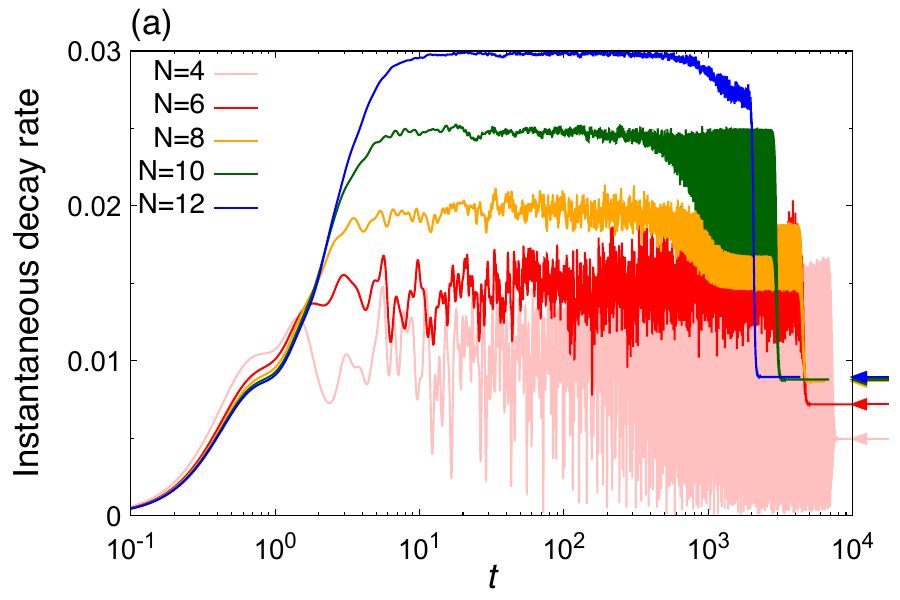}
    \includegraphics[width=0.47\linewidth]{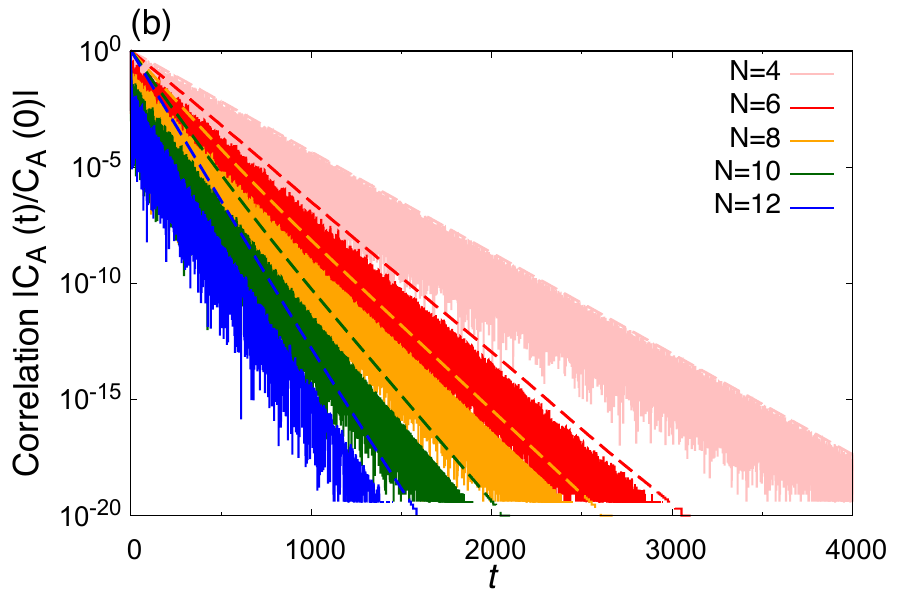}
    \caption{Dynamics of  (a) the instantaneous decay rate $\kappa_A(t)$ and (b) the autocorrelation function $C_{AA}(t)$ with $\hat{A}= \hat{n}_1 -1/2$ for various system sizes in the bulk-dissipated Bose-Hubbard model.
    The arrows in \cref{fig:BH}(a) indicate the asymptotic values of the instantaneous decay rate (i.e., the Liouvillian gap).
    The dashed lines in \Cref{fig:BH}(b) show the upper bound of the autocorrelation functions in \cref{eq:C_inst} in the main text.
    }
    \label{fig:BH}
\end{figure*}

\section{Derivation of \Cref{eq:time-derivative}}
\begin{figure*}[h]
    \flushleft
    \includegraphics[width=0.3\linewidth]{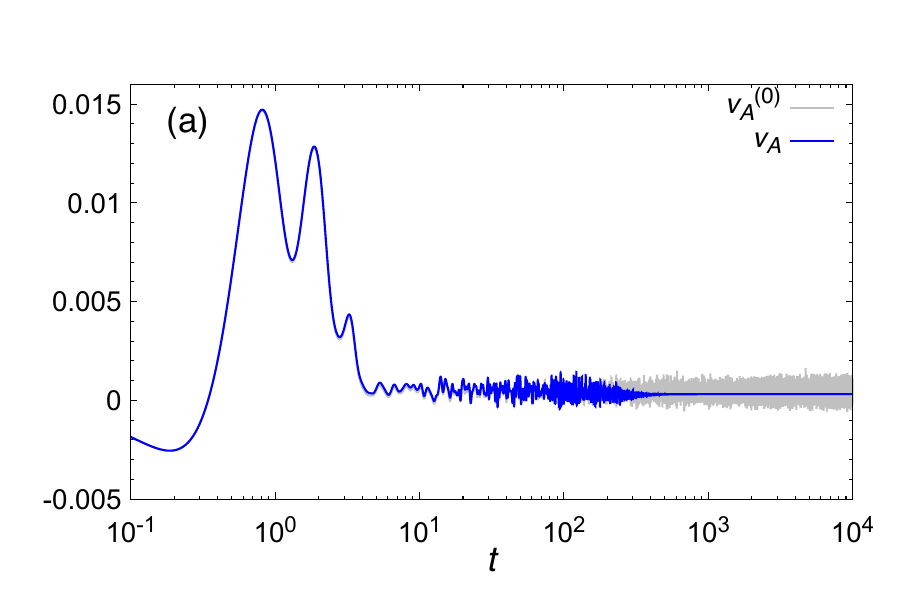}
    \includegraphics[width=0.3\linewidth]{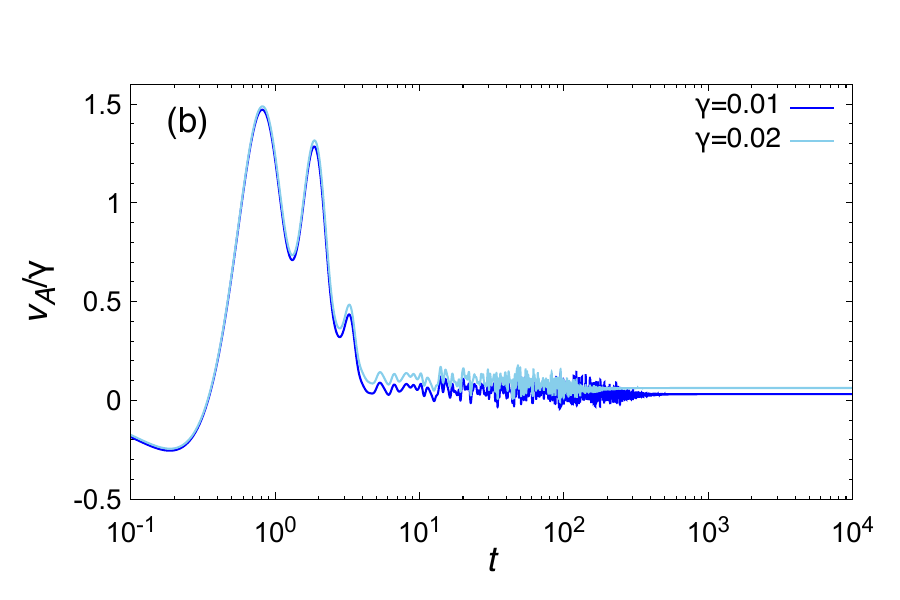}
    \includegraphics[width=0.3\linewidth]{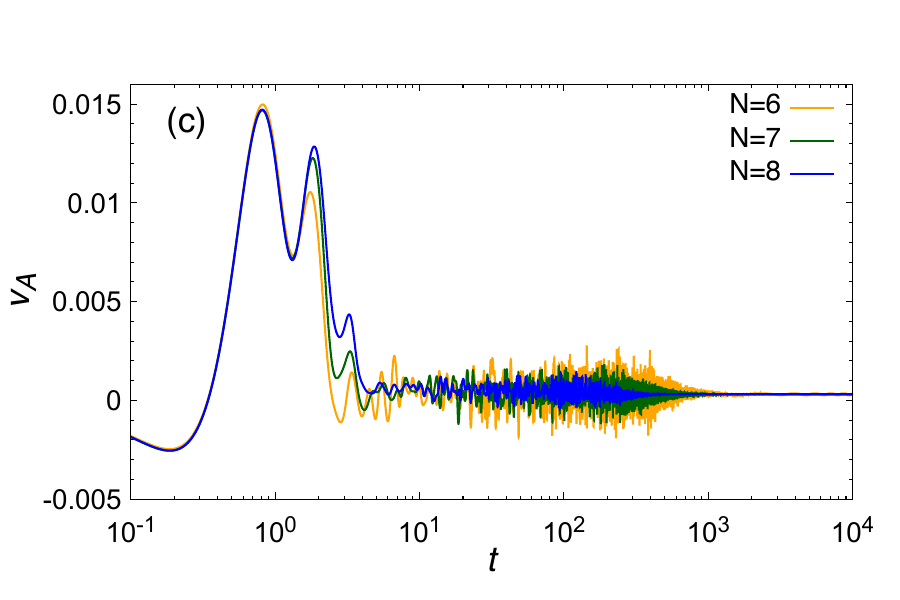}
    \includegraphics[width=0.3\linewidth]{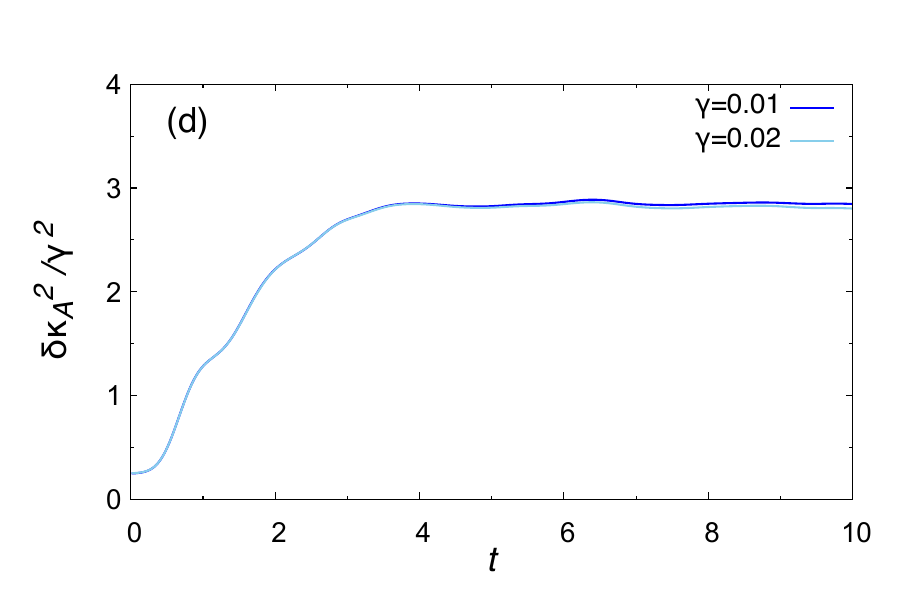}
    \includegraphics[width=0.3\linewidth]{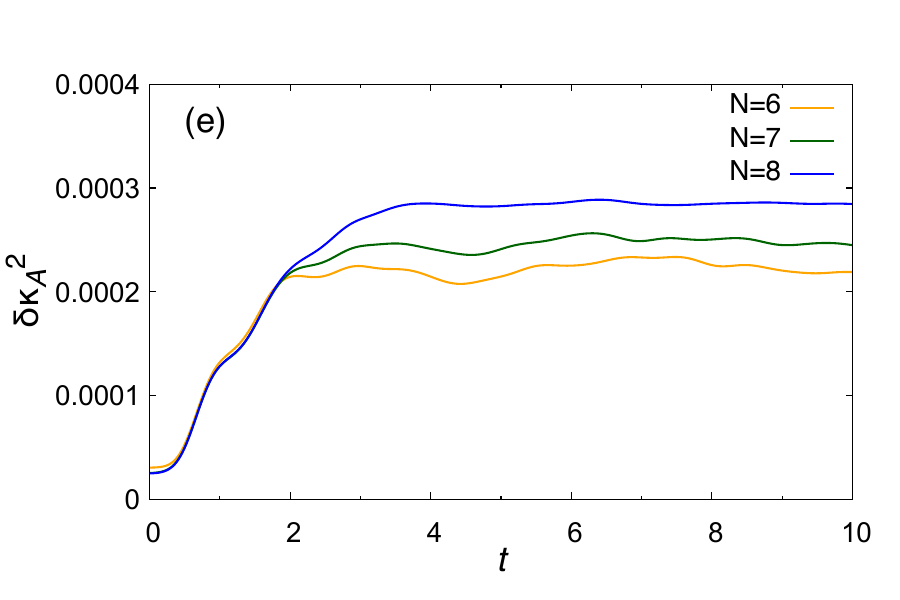}
    \includegraphics[width=0.3\linewidth]{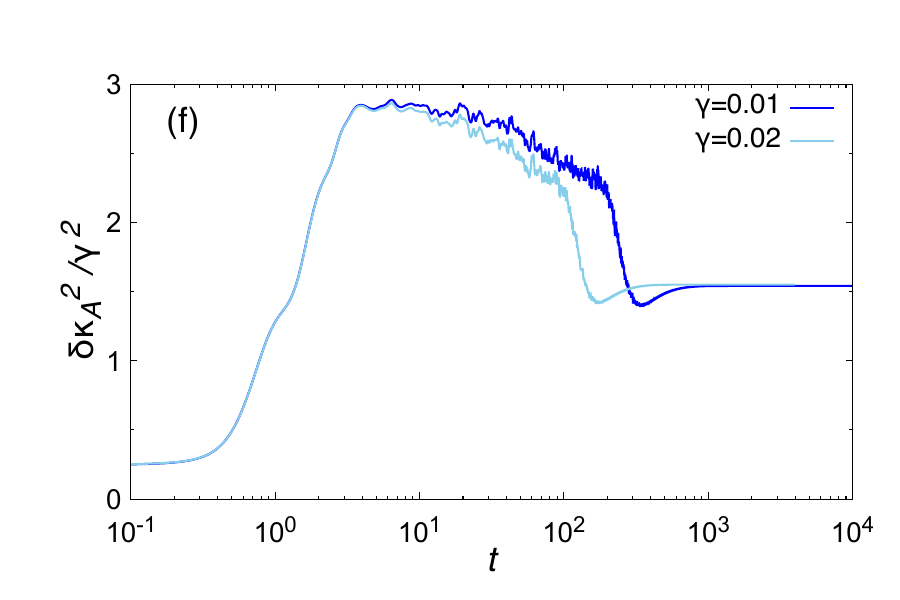}
    \includegraphics[width=0.3\linewidth]{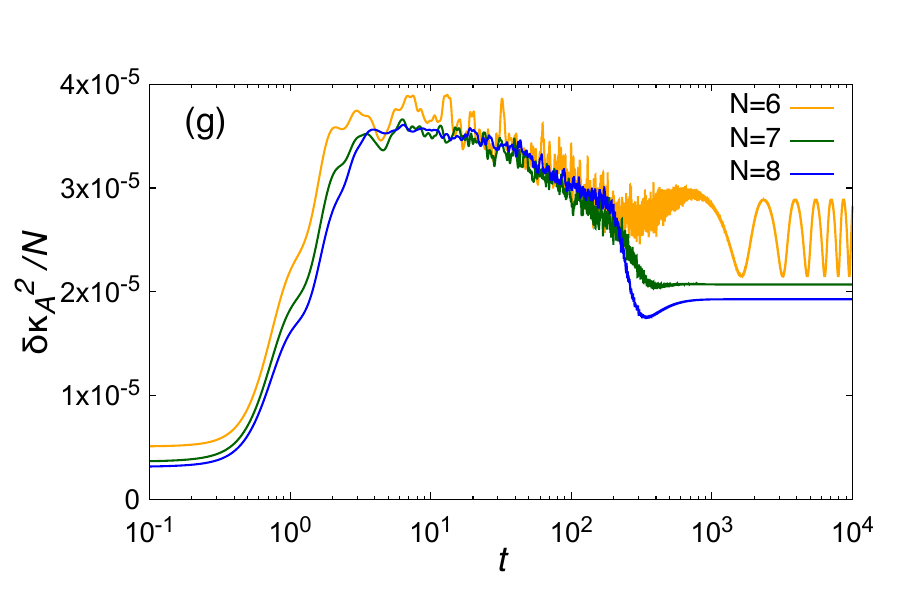}
    \caption{Dynamics of $v_A(t)$ and $\delta \kappa_A(t)^2$ with $\hat{A}=\hat{\sigma}_1^z-\braket{\hat{\sigma}_1^z}_{\rm ss}$ in the bulk-dissipated Ising spin chain (see the model in main text).
    \Cref{fig:derivative}~(a), (b), (d), and (f) use the data for $N=8$ and \cref{fig:derivative}~(a), (c), (e), and (g) use the data for $\gamma=0.01$.
    }
    \label{fig:derivative}
\end{figure*}
We derive \cref{eq:time-derivative} in the main text.
The time derivative of $\kappa_A(t)$ in \cref{eq:inst_decay} gives
\begin{align}
    \frac{\dd}{\dd t}\kappa_A(t)=& -\frac{d}{dt}\frac{\braket{\hat{A}(t), \tilde{\mathcal{L}}_s\hat{A}(t)}_\mathrm{ss}}{\braket{\hat{A}(t),\hat{A}(t)}_\mathrm{ss}}\nonumber\\
    =& -\frac{\braket{\hat{A}(t), (\tilde{\mathcal{L}}_s \tilde{\mathcal{L}}+\tilde{\mathcal{L}}^* \tilde{\mathcal{L}}_s) \hat{A}(t)}_\mathrm{ss}}{\braket{\hat{A}(t),\hat{A}(t)}_\mathrm{ss}} + \frac{2\braket{\hat{A}(t), \tilde{\mathcal{L}}_s\hat{A}(t)}_\mathrm{ss}^2}{\braket{\hat{A}(t),\hat{A}(t)}_\mathrm{ss}^2}\nonumber\\
    =&\frac{\braket{\hat{A}(t), [\tilde{\mathcal{L}}, \tilde{\mathcal{L}}_s] \hat{A}(t)}_\mathrm{ss}}{\braket{\hat{A}(t),\hat{A}(t)}_\mathrm{ss}} - 2\left( \frac{\braket{\hat{A}(t), \tilde{\mathcal{L}}_s^2 \hat{A}(t)}_\mathrm{ss}}{\braket{\hat{A}(t),\hat{A}(t)}_\mathrm{ss}} - \frac{\braket{\hat{A}(t), \tilde{\mathcal{L}}_s\hat{A}(t)}_\mathrm{ss}^2}{\braket{\hat{A}(t),\hat{A}(t)}_\mathrm{ss}^2} \right) = v_A -2\delta \kappa_A^2.
    \label{eq:dynamics_K}
\end{align}
For the first term of the last expression of \cref{eq:dynamics_K}, it is plausible to approximate $v_A$ by
\begin{equation}
    v_A^{(0)}=\frac{\braket{\hat{A}^{(0)}(t), [\tilde{\mathcal{L}}, \tilde{\mathcal{L}}_s] \hat{A}^{(0)}(t)}_\mathrm{ss}}{\braket{\hat{A}^{(0)}(t),\hat{A}^{(0)}(t)}_\mathrm{ss}} \text{  with  } \hat{A}^{(0)}(t)=e^{i\hat{H}t}\hat{A} e^{-i\hat{H}t},
\end{equation}
which is obtained by replacing $\hat{A}(t)$ in $v_A$ by $\hat{A}^{(0)}(t)$.
Indeed, our numerical calculation presented in \cref{fig:derivative}~(a) shows the validity of this approximation at arbitrary $t$ for small $\gamma$.
Thus, $v_A$ in \cref{eq:dynamics_K} expresses the rate of operator spreading due to the unitary dynamics.
In contrast, $-2\delta \kappa_A^2$ corresponds to suppression of the operator growth due to dissipation.
The quantity $\delta \kappa_A^2$ roughly gives the variance of the operator size.

In the acceleration regime, \cref{fig:derivative}~(b) and (d) imply $v_A \propto \gamma$ and $\delta \kappa_A^2 \sim \gamma^2 t$, respectively. \Cref{fig:derivative}~(c) and (e) imply both $v_A$ and $\delta \kappa_A^2$ are independent of the system size in the acceleration regime.
This result is consistent with the recent studies on random unitary circuits~\citep{keyserlingk2018operator,nahum2018operator}.
The plateau regime corresponds to the region of time in which $v_A/\gamma \approx 0$, where $\delta \kappa_A^2$ is also saturated (see \cref{fig:derivative} (e)).
\Cref{fig:derivative}~(f) and (g) show that $\delta \kappa_A^2 \sim \gamma^2 N$ in this regime.
Finally, the system reaches the asymptotic regime, where $v_A=2\delta \kappa_A^2$ and $\kappa_A(t)$ remains constant.

\section{Dissipation strength dependence of instantaneous decay rate}
\begin{figure}[h]
    \flushleft
    \includegraphics[width=0.3\linewidth]{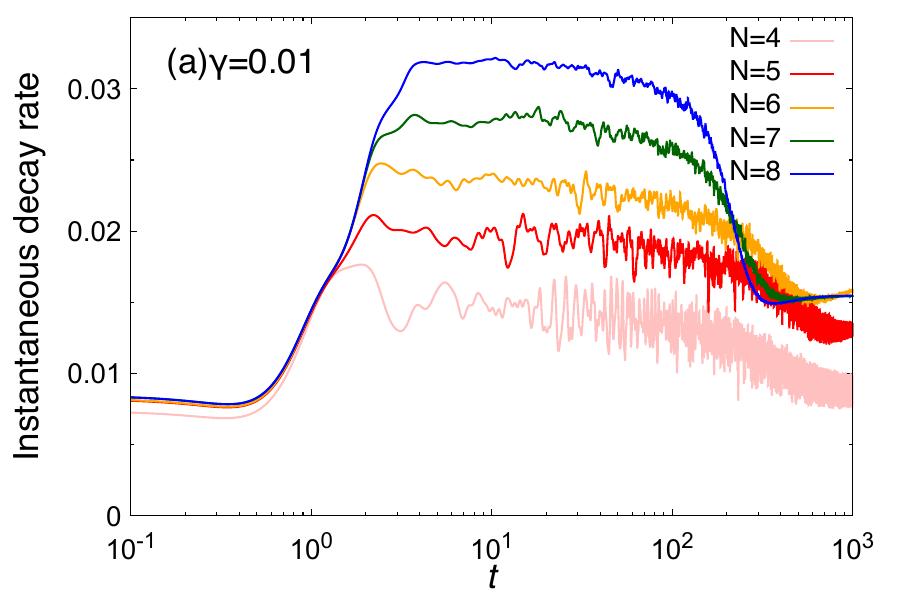}
    \includegraphics[width=0.3\linewidth]{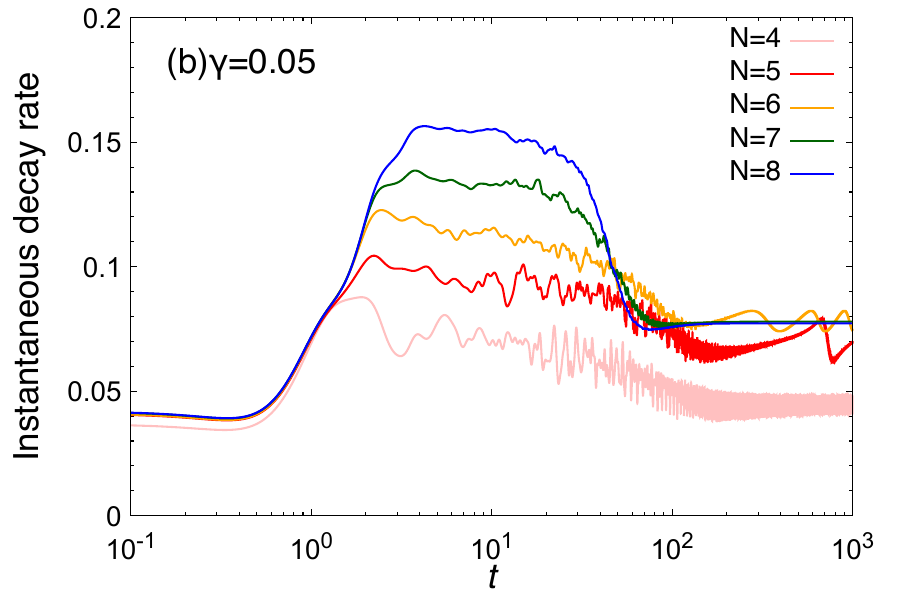}
    \includegraphics[width=0.3\linewidth]{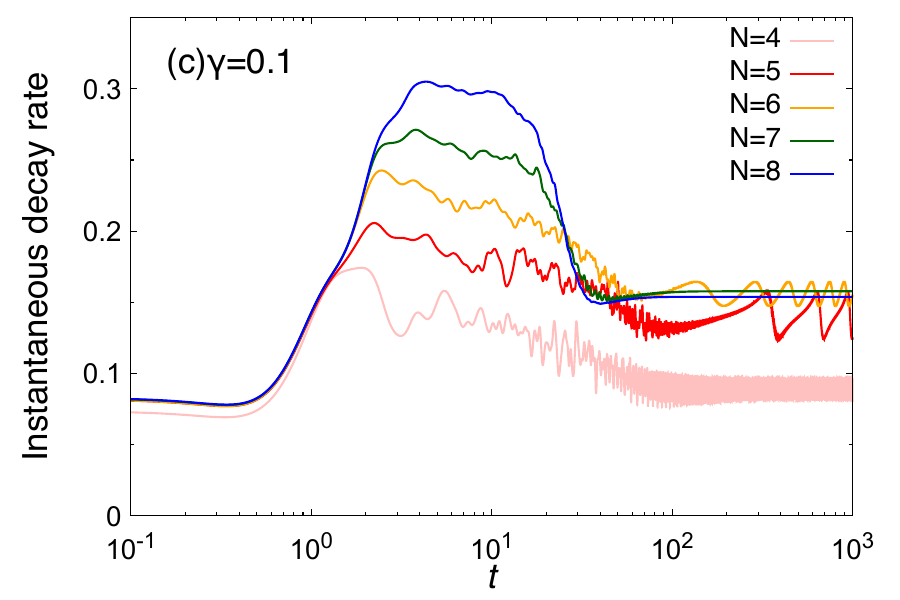}
    \includegraphics[width=0.3\linewidth]{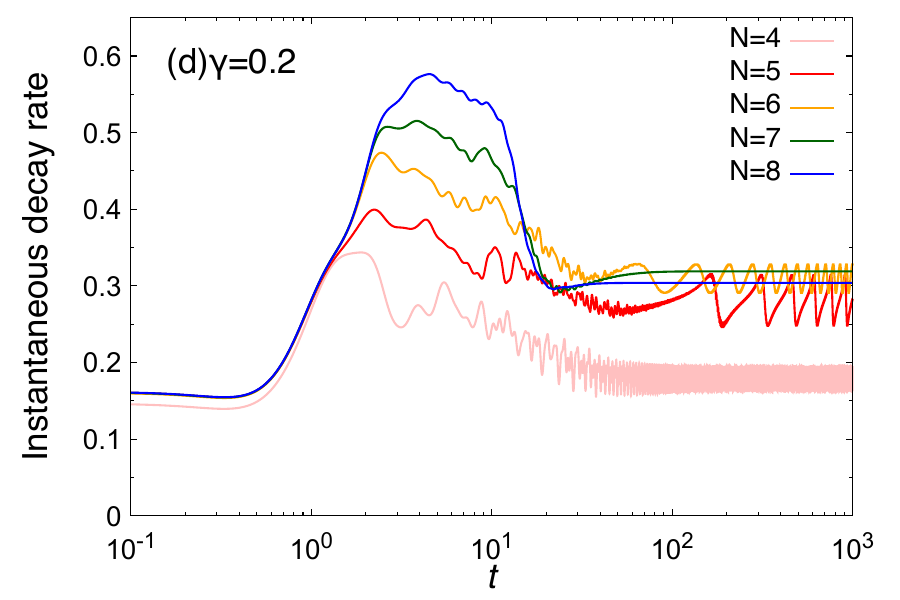}
    \includegraphics[width=0.3\linewidth]{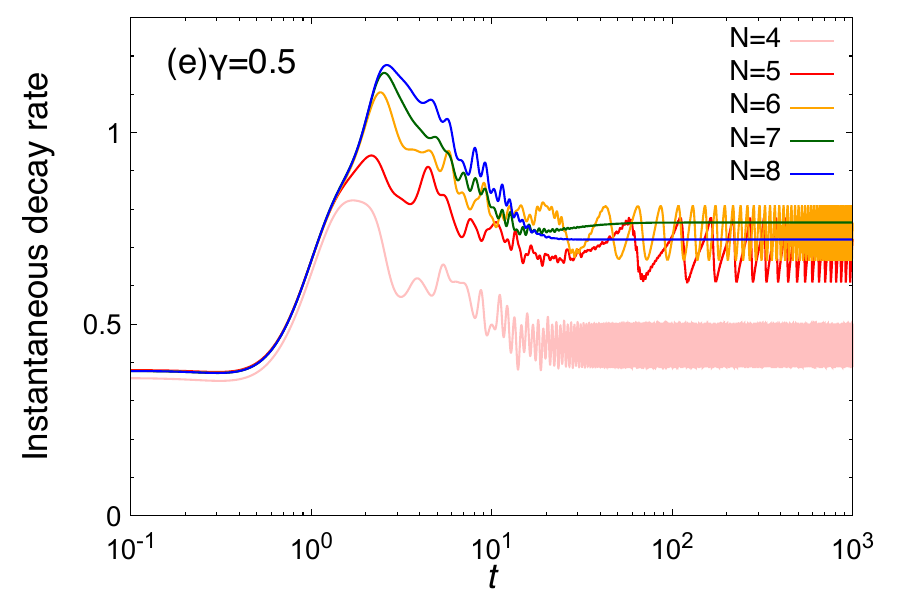}
    \includegraphics[width=0.3\linewidth]{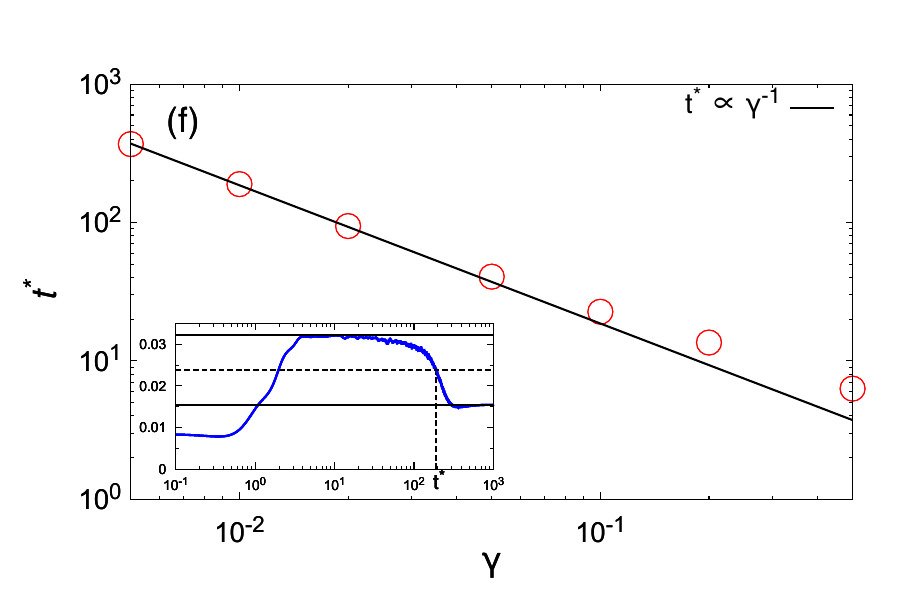}
    \includegraphics[width=0.3\linewidth]{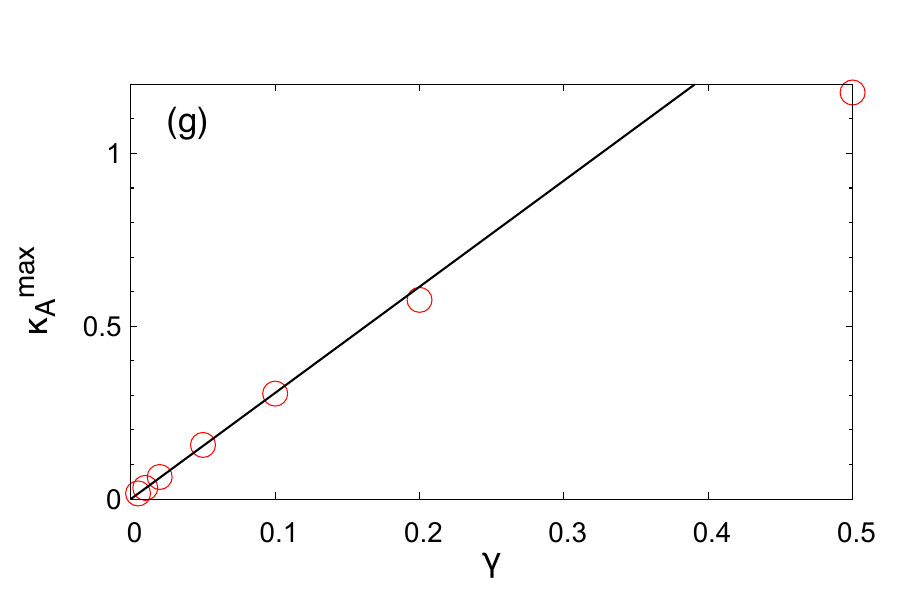}
    \includegraphics[width=0.3\linewidth]{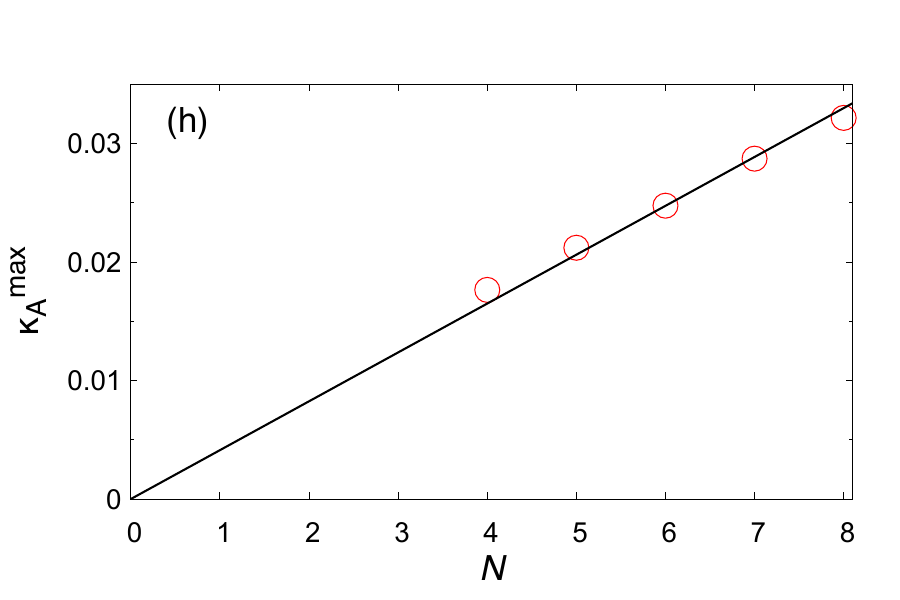}
    \caption{(a)--(e): Dynamics of the instantaneous decay rate $\kappa_A (t)$ with $\hat{A}=\hat{\sigma}_1^z - \braket{\hat{\sigma}_1^z}_{\mathrm{ss}}$ for various dissipation strength $\gamma$ in the bulk-dissipated Ising spin chain.
    (f): $\gamma$-dependence of the crossover time into the asymptotic regime $t^*$. 
    Line is a guide to show the scaling of $t^*$ for weak dissipation.
    Inset describes how to determine $t^*$.
    (g): $\gamma$-dependence of the maximum value of the instantaneous decay rate $\kappa_A^{\mathrm{max}}$ at $N=8$.
    Line shows $\kappa_A^{\mathrm{max}} \propto \gamma$ for weak dissipation.
    (h): $N$-dependence of $\kappa_A^{\mathrm{max}}$ at $\gamma=0.01$.
    Line shows $\kappa_A^{\mathrm{max}} \propto N$.
    }
    \label{fig:gamma_dependence}
\end{figure}

\Cref{fig:gamma_dependence}~(a)--(e) show the dynamics of the instantaneous decay rate for different dissipation strengths $\gamma$ in the spin-chain model, as discussed in the main text.
As the dissipation increases from \cref{fig:gamma_dependence}~(a) to (e), the plateau regime is narrower.
When $\gamma = 0.5$, the plateau regime disappears, and the peak of $\kappa_A (t)$ almost saturates at $N=7$.
\Cref{fig:gamma_dependence}~(f) shows $\gamma$-dependence of the crossover time $t^*$ into the asymptotic regime.
The inset of figure describes how to determine $t^*$.
The horizontal solid lines represent the maximum of the instantaneous decay rate $\kappa_A^{\mathrm{max}}$ and $g$, while the horizontal dotted line corresponds to $\kappa_A^{\mathrm{med}}$, which is the average of $\kappa_A^{\mathrm{max}}$ and $g$.
The crossover time $t^*$ is defined as the time when $\kappa_A(t)$ intersects with $\kappa_A^{\mathrm{med}}$ between the plateau regime and the asymptotic regime.
The figure implies that the crossover time is inversely proportional to $\gamma$ for weak dissipation.
While our theory predicts that $t^*$ is independent of system size, the limited system size poses challenges to its verification.
In \cref{fig:gamma_dependence}~(g), the $\gamma$-dependence of $\kappa_A^{\mathrm{max}}$ indicates $\kappa_A^{\mathrm{max}} \propto \gamma$ for weak dissipation.
\Cref{fig:gamma_dependence}~(h) also shows $\kappa_A^\mathrm{max}\propto N$.
Thus, our theoretical prediction that $\kappa_A \propto N \gamma$ in the plateau regime is fully verified in numerics.

\clearpage
\end{widetext}

\end{document}